\documentclass[journal]{IEEEtran}

\usepackage{cite}

\ifCLASSINFOpdf
  \usepackage[pdftex]{graphicx}
\else
  \usepackage[dvips]{graphicx}
\fi

\usepackage{amsmath}
\usepackage{amsfonts}
\usepackage{array} 
\usepackage{tikz}
\usepackage{xcolor}
\usepackage[inline]{enumitem}
\usepackage{tabularx}
 
\usepackage[most]{tcolorbox}
\usepackage{threeparttable}
\usepackage{booktabs}
\PassOptionsToPackage{hyphens}{url}
\usepackage{hyperref}
\usepackage{makecell}
\usepackage{wasysym}

\newtheorem{definition}{Definition}[section]

\newtcolorbox[%
auto counter]{mybox}[2][]{%
	enhanced jigsaw,
        colback=yellow!12,
	breakable,
	#1}
\begin{document}
\title{On Protecting the Data Privacy of Large Language Models (LLMs): A Survey}

\author{
	Biwei Yan,
        Kun Li,
        Minghui Xu,
        Yueyan Dong,
        Yue Zhang,
        Zhaochun Ren,
        Xiuzhen Cheng

	\IEEEcompsocitemizethanks{
            \IEEEcompsocthanksitem B. Yan, K. Li, M. Xu, Y. Dong, X. Cheng are with the School of Computer and Science and Technology, Shandong University. Email: \{bwyan, kli, mhxu, xzcheng\}@sdu.edu.cn\\
            \IEEEcompsocthanksitem Z. Ren is with Leiden University. Email: \{z.ren@liacs.leidenuniv.nl\}\\
            \IEEEcompsocthanksitem Y. Zhang is with the Department of Computer Science, Drexel University. Email: \{yz899@drexel.edu\}\\
		% \IEEEcompsocthanksitem
	}

}

\maketitle

\begin{abstract}
Large language models (LLMs) are complex artificial intelligence systems capable of understanding, generating and translating human language. They learn language patterns by analyzing large amounts of text data, allowing them to perform writing, conversation, summarizing and other language tasks. When LLMs process and generate large amounts of data, there is a risk of leaking sensitive information, which may threaten data privacy. This paper concentrates on elucidating the data privacy concerns associated with LLMs to foster a comprehensive understanding. Specifically, a thorough investigation is undertaken to delineate the spectrum of data privacy threats, encompassing both passive privacy leakage and active privacy attacks within LLMs. Subsequently, we conduct an assessment of the privacy protection mechanisms employed by LLMs at various stages, followed by a detailed examination of their efficacy and constraints. Finally, the discourse extends to delineate the challenges encountered and outline prospective directions for advancement in the realm of LLM privacy protection.
\end{abstract}

% \fi
\IEEEpeerreviewmaketitle

\section{Introduction}

% \ZY{Regarding the title, should we highlight data privacy? or privacy is fine...}

In recent years, Large Language Models (LLMs) have emerged as pivotal players in the realm of artificial intelligence, revolutionizing various fields such as natural language processing \cite{gao2024llm, xie2023translating}, embodied AI \cite{song2023llm, xu2024survey, duan2022survey}, AI-generated content (AIGC) \cite{cao2023comprehensive, wu2023ai}. LLMs, trained on massive datasets, possess the remarkable ability to generate human-like text, answer complex queries, and undertake a myriad of language-related tasks with unprecedented accuracy and fluency. However, amidst the excitement surrounding the capabilities of LLMs, concerns about data privacy have garnered increasing attention \cite{yao2023survey}. \looseness=-1

On one hand, LLMs may be subject to passive privacy leakage. 
Users can inadvertently expose sensitive data to ChatGPT if they input such information into the chat interface. For example, Samsung Electronics experienced inadvertent leakage of sensitive company data through ChatGPT in three distinct occurrences. 
% In the initial incident, an engineer transferred flawed source code retrieved from a semiconductor database to ChatGPT for error correction. Subsequently, an employee utilized ChatGPT to enhance code aimed at detecting defects in Samsung equipment. Additionally, another employee sought ChatGPT's assistance in crafting minutes for an internal meeting at Samsung, collectively contributing to the occurrences of data leaks. 
Besides, LLMs often rely on vast amounts of data for training, including text scraped from the internet, publicly available datasets, or proprietary sources. This data aggregation process can raise significant data privacy concerns, especially when dealing with sensitive or personally identifiable information (PII) \cite{subramani2023detecting}. LLMs have been shown to have the potential for memorization of training data, raising concerns about inadvertent leakage of sensitive information during inference \cite{staab2023beyond}. Even with techniques such as differential privacy or federated learning, which aim to mitigate privacy risks during training, residual traces of sensitive data may still persist within the model's parameters \cite{zhao2023privacy}. 
% \ZY{You can add an example here to demonstrate the impact of passive privacy issues in LLMs.}
%As these models continue to grow in sophistication and deployment, understanding the nuances of safeguarding sensitive information within them becomes paramount.

On the other hand, LLMs may be vulnerable to active privacy attacks. The deployment of fine-tuned LLMs in various applications introduces additional security challenges. Fine-tuning or adapting pre-trained LLMs to specific tasks may inadvertently expose them to the exploitation of vulnerabilities, potentially compromising the confidentiality, integrity, or availability of sensitive information \cite{luo2023taiyi}. For example, to bypass the model's inherent alignment, a prompting strategy was devised that induces GPT-3.5-turbo to ``diverge" from producing conventional responses, instead emit training data \cite{nasr2023scalable}. Pre-existing vulnerabilities such as backdoor attacks, membership inference attacks, and model inversion attacks can be leveraged against pre-trained or fine-tuned models with the objective of illicitly acquiring sensitive data.

% \ZY{You can add an example here to demonstrate the impact of active privacy attacks in LLMs.}
% \ZY{I assume the following is a part of passive privacy leakage, no? If so, please somehow integrate it with the previous paragraphs}
%

\begin{figure*}[!htb]
\centering
\centerline{\includegraphics[width=1\linewidth]{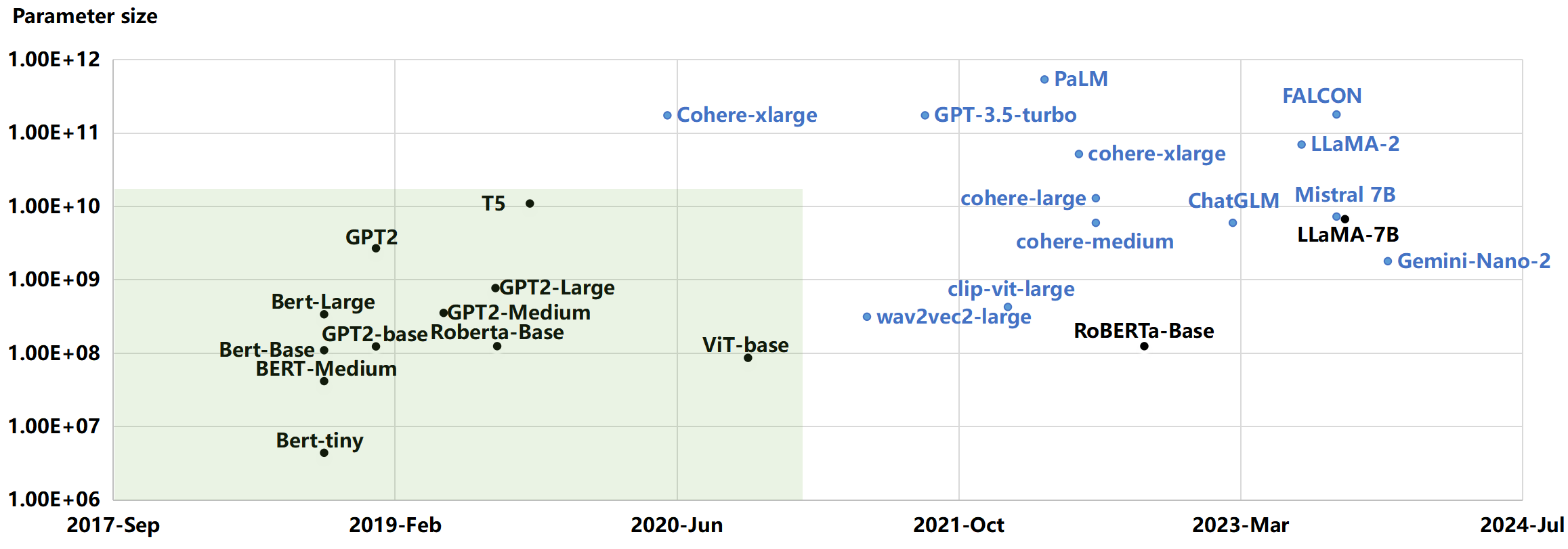}}
\caption{The current state of research on privacy protection for LLMs is depicted. The horizontal axis represents the time of LLM releases, while the vertical axis represents the size of model parameters. Blue dots signify LLM instances not addressed in literature pertaining to privacy protection, whereas black dots indicate those that have been examined in such literature. The green backdrop delineates the central cluster zone of LLMs with the potential to facilitate privacy protection.}
\label{fig:overview}
\end{figure*}

% \ZY{
% The figure is good, but it's not clear how you created it. You may need to discuss that at a higher level. Also, I think it may be included in Section IV to further motivate the paper (you decide).}
To portray the current situation, we outline the present state of research concerning privacy safeguards for LLMs in Fig. \ref{fig:overview}. 
Taking into account academic papers on privacy protection and the model list from Hugging Face, we have compiled a list of popular LLMs in the figure. The timeline axis represents the release dates of models, while the vertical axis indicates the size of parameters. Blue data points signify LLMs that have received limited attention in the literature regarding privacy protection, while black data points indicate models studied alongside privacy safeguards. Currently, scholarly focus on data privacy in LLMs primarily revolves around well-known models of relatively smaller scale, like pre-2020 versions of the GPT-2 \cite{radford2019language} and BERT \cite{devlin2018bert} series. In contrast, recent releases of LLMs with larger parameter sizes have not been adequately scrutinized due to some models not being publicly available, and privacy protection technology lagging behind the rapid development of LLMs. 

In this paper, we extensively investigate data privacy concerns within Large LLMs, specifically examining potential privacy threats from two folds: privacy leakage and privacy attacks. Besides, we delve into the corresponding countermeasures by providing a comprehensive review from the three major stages of developing LLMs: pre-training, fine-tuning and inference. Our contributions are summarized as follows:

% \ZY{The introduction appears brief and lacks a thorough discussion of major methods of privacy protection (your title is privacy protection for LLM, but you do not discuss that in the intro, which is weird). Expanding on this topic is crucial as it forms the core of the paper. It's essential to delve into various privacy protection techniques, summarizing key findings as necessary to provide a comprehensive overview. }

\begin{itemize}
\item We undertook a comprehensive investigation into the scholarly literature concerning privacy threats within LLMs, categorizing them into two distinct groups: privacy leakage and privacy attacks.

\item Our examination encompasses an analysis of privacy protection methodologies applied to LLMs, which we categorized based on developmental stages. We categorize privacy protections into three groups based on their location: pre-training, fine-tuning, and inferences. Within each category, we introduce techniques at a high level, explain their application in LLMs, and provide a detailed literature review. 
The goal of our survey is to provide guidance for LLM developers on implementing cutting-edge techniques to safeguard LLMs.

%Our survey aims to offer guidelines to instruct LLM developers on how to protect LLMs using state-of-the-art techniques.

%\item Through an exhaustive review of existing literature, we have synthesized fresh perspectives on data privacy issues within LLMs, thereby offering substantial direction for future research endeavors.
\end{itemize}

\section{Related Work} 
In this section, we first introduce existing surveys about the development and evaluation of LLMs. Then, we further elaborate the most related work addressing the privacy and security issues in LLMs, and finally summarize the research of our survey.

\subsection{Surveys on LLM Evaluation}
Currently, some works have surveyed the development and evaluation of LLMs. These studies typically cover architectural improvements of LLMs (such as the GPT series, BERT, Transformers~ \cite{li2024personal,chang2023survey,zhao2023survey, wu2023survey,  bowman2023eight, naveed2023comprehensive,hadi2023survey}). For example, Li \textit{et al.} \cite{li2024personal} focused on integrating LLM with intelligent personal assistants (IPAs) to improve personal assistance capabilities. It delves into the architecture, capabilities, efficiency, and security aspects of these agents. Zhao \textit{et al.} \cite{zhao2023survey} focused on four key aspects of LLMs: pre-training, adaptation tuning, utilization, and capacity evaluation. It provides a thorough background on LLMs, including terminologies and techniques. Naveed \textit{et al.} \cite{naveed2023comprehensive} provided an extensive analysis of LLMs, covering their architecture, training, applications, and challenges. It dives into detailed aspects of LLMs like pre-training, fine-tuning, and evaluation, while also discussing various LLM applications in different fields. Hadi \textit{et al.} \cite{hadi2023survey} introduced a thorough overview of LLMs, discussing their history, training, and applications in various fields like medicine, education, finance, and engineering. It examines the technical aspects, challenges, and future potential of LLMs, including ethical considerations and computational requirements.

% \ZY{I am not sure whether we should include the measurements of LLMs?}
To understand the capabilities and limitations of LLMs in various applications, some works have conducted comprehensive measurements on these LLMs \cite{chang2023survey,guo2023evaluating,liu2023trustworthy}. Chang \textit{et al.} \cite{chang2023survey} offered a comprehensive analysis of the methods and criteria for evaluating LLMs. It discusses various aspects including tasks to evaluate, datasets, benchmarks, and evaluation techniques. Guo \textit{et al.} \cite{guo2023evaluating} emphasized the need for a comprehensive evaluation of LLMs in various dimensions, such as knowledge and capability evaluation, alignment evaluation, security considerations, and applications in the specialized domain. In \cite{liu2023trustworthy}, Liu \textit{et al.} examined the alignment of LLMs with human values and social norms. It proposes a detailed taxonomy to evaluate LLM trustworthiness on various dimensions such as reliability, safety, fairness, resistance to misuse, explainability, adherence to social norms, and robustness.

\subsection{Surveys on LLM Security and Privacy}
Since the training of LLMs relies on a substantial amount of data, which usually includes sensitive information. Therefore, LLMs face challenges in handling privacy and security issues \cite{yao2023survey,li2023privacy,neel2023privacy,marshall2023effects,al2023chatgpt,qammar2023chatbots,schwinn2023adversarial,derner2023beyond,shayegani2023survey}. Yao \textit{et al.} \cite{yao2023survey} comprehensively investigated the security and privacy of LLMs, and conducted an extensive review of the literature on LLMs from three aspects: beneficial security applications (such as vulnerability detection, secure code generation), adverse effects (e.g., phishing attacks, social engineering) and vulnerabilities (e.g., jailbreak attacks, prompt attacks), as well as corresponding defense measures. Li \textit{et al.} \cite{li2023privacy} delved into privacy concerns in LLMs, categorizing privacy attacks and detailing defense strategies. It also explores future research directions for enhancing privacy in LLMs. Neel \textit{et al.} \cite{neel2023privacy} explored the privacy risks associated with LLMs, focusing on issues such as the memory of sensitive data and various privacy attacks. It reviews mitigation techniques and highlights the current state of privacy research in LLMs. However, they mainly focus on work that red-teams models to highlight privacy attacks. 

Marshall \textit{et al.} \cite{marshall2023effects} and Al-Hawawreh \textit{et al.} \cite{al2023chatgpt} explored the role of ChatGPT in the field of cybersecurity. Their discussions emphasized its real-world uses, such as enhancing code security and detecting malware. Qammar \textit{et al.} \cite{qammar2023chatbots} provided an extensive overview of the evolution of chatbots to ChatGPT and their role in cybersecurity, highlighting vulnerabilities and potential attacks. However, it may lack depth in specific cybersecurity solutions and preventive measures against identified vulnerabilities and attacks. Schwinn \textit{et al.} \cite{schwinn2023adversarial} offered a comprehensive analysis of both old and new threats in LLMs, providing insight into evolving adversarial attacks and defenses. But The focus on a broad range of threats might overlook in-depth details on specific attack methodologies or defense mechanisms. Derner \textit{et al.} \cite{derner2023beyond} investigated specific security risks associated with ChatGPT, contributing to a better understanding of its vulnerabilities. However, it may not provide a comprehensive comparison with other models or systems, limiting its scope to ChatGPT only. Shayegani \textit{et al.} \cite{shayegani2023survey} thoroughly examined the vulnerabilities in LLMs exposed by adversarial attacks, offering valuable insights for future model improvements. Nonetheless, the focus on adversarial attacks might lead to less emphasis on other types of vulnerabilities or broader security issues.

% \ZY{our primary contribution, in comparison to another survey, lies in our comprehensive literature review on privacy protections? } 

In contrast to existing surveys, our research concentrates on addressing data privacy issues within LLMs, providing a comprehensive literature reviwe of privacy threats and privacy protection techniques. We thoroughly examine the countermeasures employed to mitigate privacy threats at different stages, and engage in an in-depth discussion on the current challenges and future research directions in LLM data privacy, aiming to offer guidance and reference for this field.

\section{Background on Large Language Models (LLMs)}\label{sec:preliminary}
%To obtain a holistic comprehension of the diverse methodologies employed in safeguarding the privacy of LLMs, we present the pertinent concepts encompassing LLMs and cryptographic tools.

LLMs are super-large deep learning models pre-trained on vast amounts of data, containing tens of billions to trillions of parameters. They construct extensive unsupervised training based on these parameters, enabling them to more accurately learn patterns and structures of natural language, thereby understanding and generating natural language texts. Compared to traditional NLP models, LLMs demonstrate better proficiency in understanding and generating natural texts, and also exhibit certain logical thinking and reasoning abilities, which is widely in programming \cite{cai2023low}, vulnerability detection \cite{karpinska2023large}, and medical text analysis \cite{thirunavukarasu2023large}. In 2017, Vaswani \textit{et al.}\cite{vaswani2017attention} introduced the Transformer architecture, which uses parallel processing and attention mechanisms to provide an effective method for processing sequential data (especially text). This significantly enhances the efficiency of dealing with sequential data and supports more efficient training on large datasets, fostering the rapid development of LLMs such as the GPT series, BERT, and Transformer models. The training of LLMs primarily includes two key stages: pre-training and fine-tuning.

\begin{figure*}[!ht]
\centering
\centerline{\includegraphics[width=0.8\linewidth]{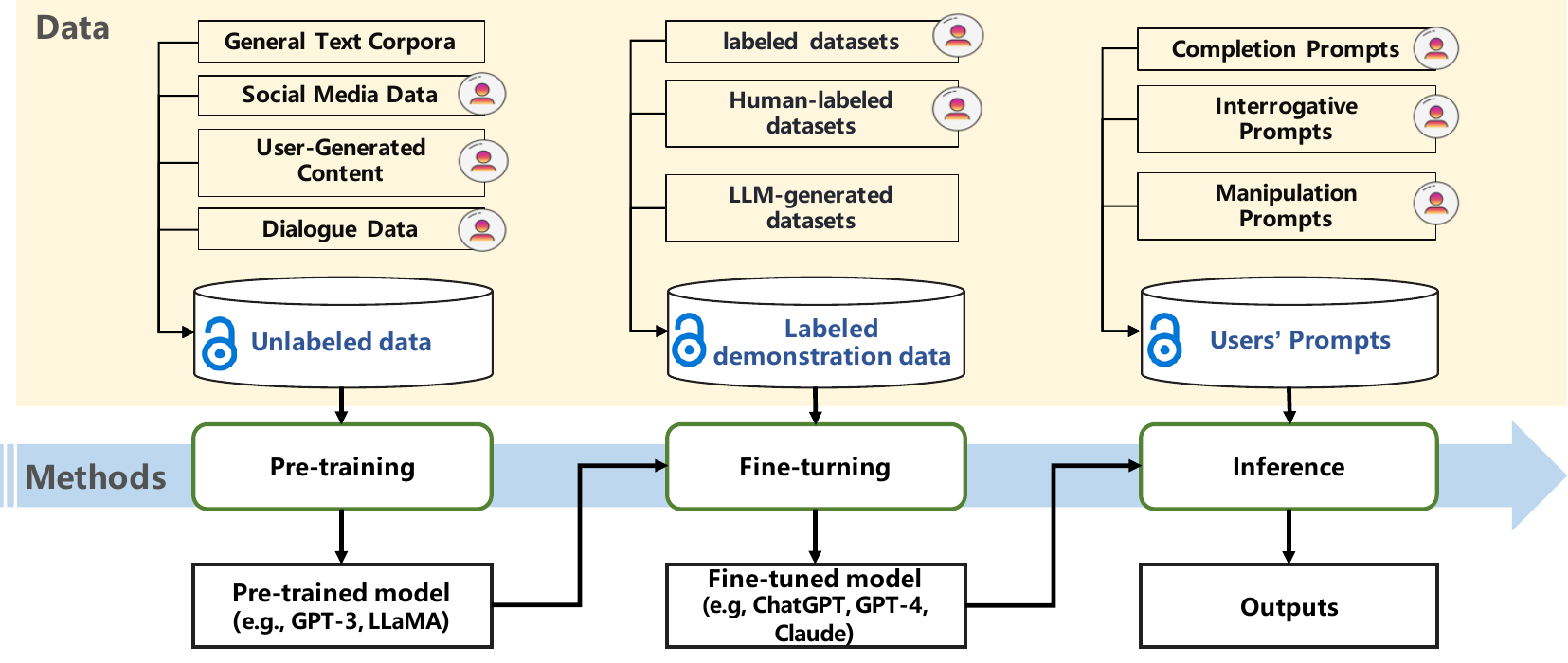}}
\caption{The process of data propagation during both the training and inference stages of LLMs.}
% \ZY{The user icon here looks like an attacker (hacker) icon}}
\label{fig:data-propagation}
\end{figure*}

\begin{itemize}
    \item \textbf{Pre-training}: At this stage, the model is typically trained on a very large and diverse dataset. These datasets may include texts from a variety of sources such as the Internet, books and news, or large text datasets published by many organizations and research institutions for academic research. E.g. general text corpora, social media data, user-generated content, and dialogue data). For example, GPT-3, developed by OpenAI, was pre-trained using
    CommonCrawl, constituting 45TB of compressed plaintext before filtering\cite{brown2020language}. Regarding multimodal LLMs, CLIP's training dataset encompasses 400 million pairs of images and text, while Stable Diffusion was trained on a dataset consisting of two billion examples sourced from LAION-2B \cite{gadre2024datacomp}.
    The purpose of pre-training is to enable the model to learn a wide range of language patterns, structures, and knowledge. Through this process, the model acquires a broad ability to understand language, including understanding vocabulary, grammar, and even some common sense. This stage does not focus on any specific task but rather provides a general foundation for language understanding.

    \item \textbf{Fine-tuning}: The fine-tuning stage is carried out on the basis of a pre-trained model, with the goal of better adapting the model to specific tasks or domains. During this phase, the model is trained on a smaller, more specific dataset that is closely related to the target task or domain. The datasets are usually sourced from websites and forums of specific professional fields, such as the medical, legal, technological, and other professional communities, and mainly consist of labeled demonstration data such as labeled datasets, human-labeled datasets, and LLM-generated datasets. The datasets available for fine-tuning may be relatively small, typically ranging from a few hundred to a few thousand text samples. Through fine-tuning, the model learns the characteristics and details specific to the task.
\end{itemize}

The advantage of this two-stage training method is that it combines the breadth of general language understanding (through pre-training) with the depth of adaptability to specific tasks (through fine-tuning). This enables the model to exhibit higher accuracy and efficiency when dealing with a variety of complex, domain-specific tasks. After the model has been trained and fine-tuned, the inference stage can be performed.

\begin{itemize}
    \item \textbf{Inference}: In this phase, the trained model is used to make prediction or decision. This includes processing input data (such as users' prompts), using the model to compute outputs, and possibly post-processing to fit specific application needs. The primary purpose of inference is to leverage the knowledge learned by the model to solve real-world problems, such as automated responses, image recognition, or other forms of data analysis.
\end{itemize}

\section{Scope, Methodology, and Overview}

%In the following, we outline the scope of this survey, the methods employed for data collection and analysis, as well as the structure of the paper.

\subsection{Scope}

Our paper is dedicated to conducting a comprehensive literature review in the field of data privacy for LLMs, organizing and reviewing existing research. We conduct a comprehensive and in-depth privacy analysis, including privacy leakage and privacy attacks in LLMs, as well as privacy protection methods at different stages of privacy inference within LLMs. Our focus is not only on the implementation details of these technologies but also on a deep exploration of their effectiveness in protecting privacy, as well as their potential limitations.
%Our goal is to thoroughly explore and assess the scenarios of privacy leakage that may occur at these different stages, providing a comprehensive perspective to better understand the specific challenges faced in achieving privacy protection in LLMs.

\begin{figure}[!htb]
\centering
\centerline{\includegraphics[width=0.9\linewidth]{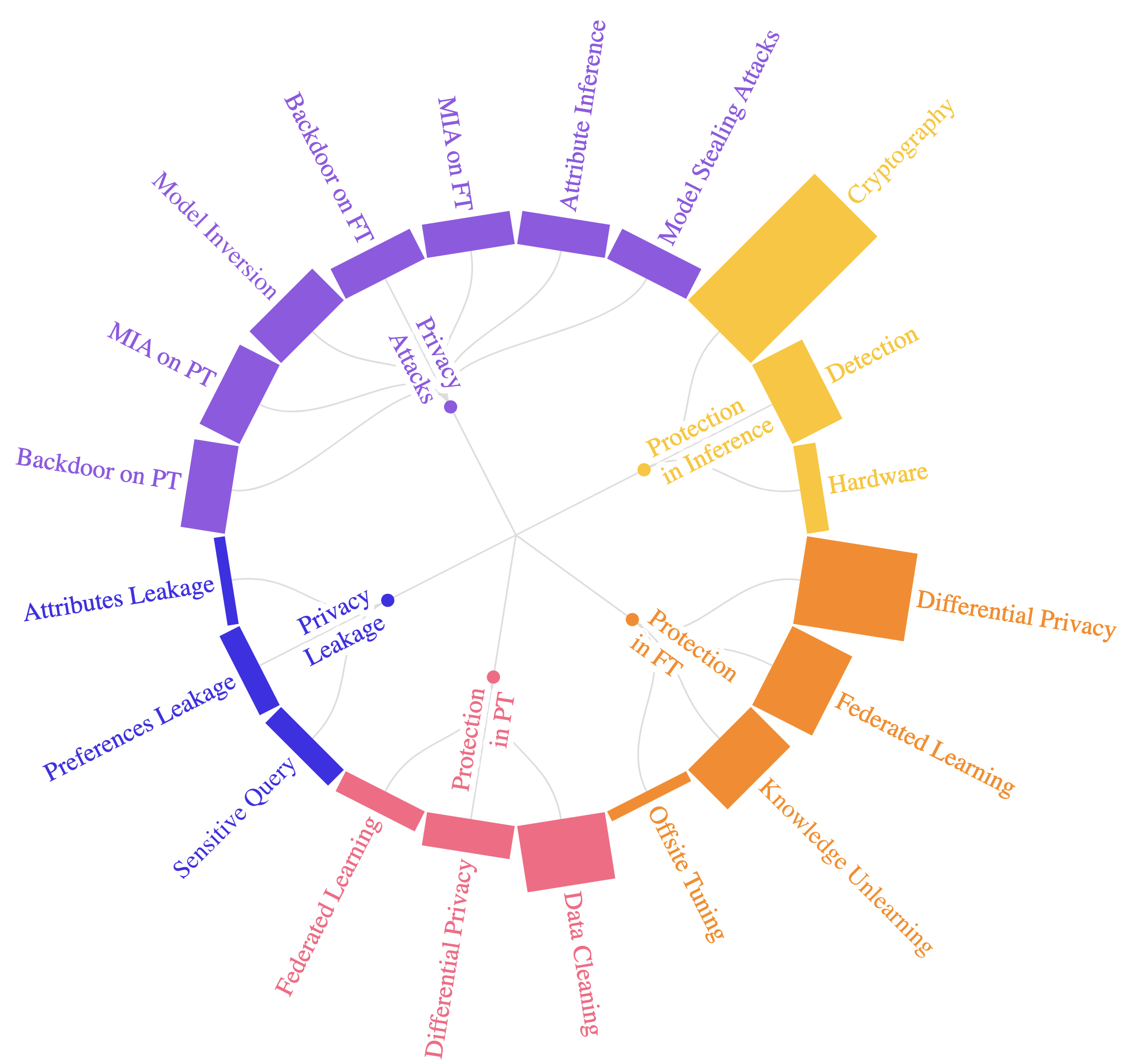}}
\caption{The distribution of research papers concerning the data privacy in LLMs. ``PT" and ``FT" represent abbreviations for Pre-Training and Fine-Tuning, respectively.}
\label{fig:data-collection} 
\end{figure}
\begin{figure*}[!htb]
\centering
\centerline{\includegraphics[width=1\linewidth]{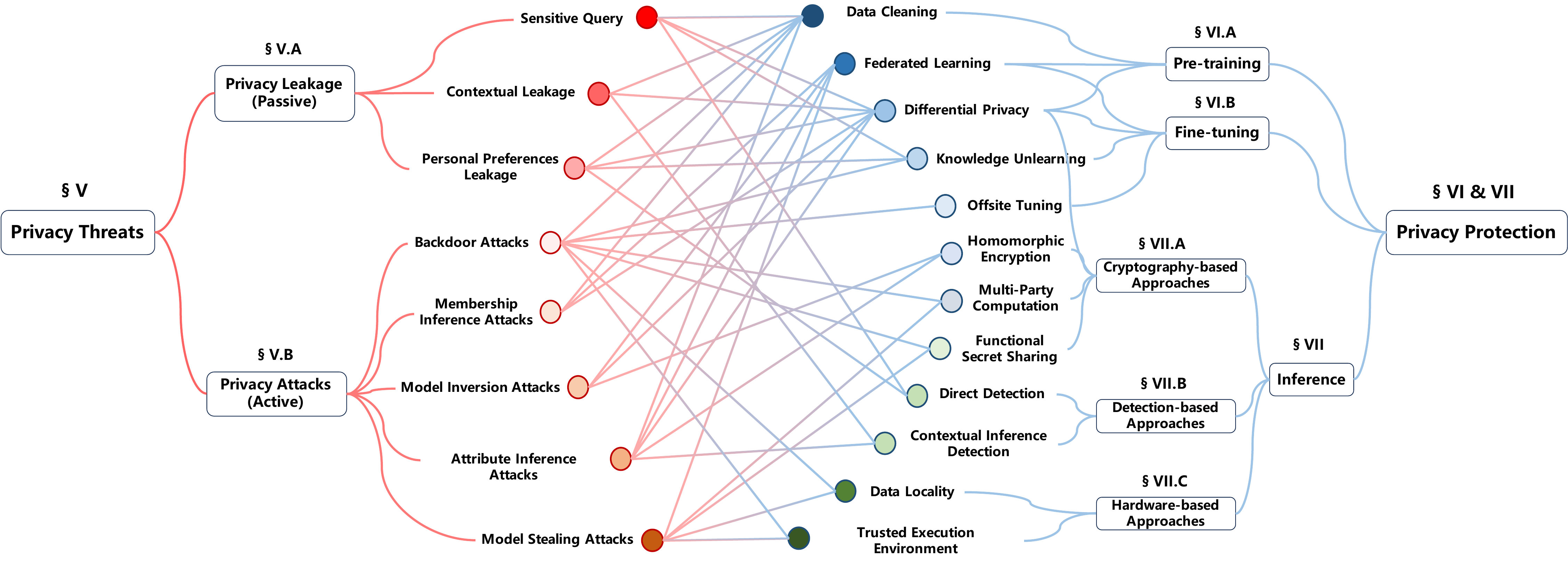}}
\caption{Privacy threats, protection, and their defensive correlations.}
\label{fig:privacy-threats}
\end{figure*}

\subsection{Methodology}

\noindent\textbf{Data Collection:} To comprehensively understand the landscape of data privacy concerns in LLMs, we executed a structured literature search on Google Scholar. The results are summarized in Fig. \ref{fig:data-collection}, wherein we categorized the retrieved literature into distinct themes. From the 91 collected papers, we identified 33 that specifically highlight the privacy threats confronting LLMs. Within this subset, a division reveals that 5 papers focus on privacy leakage, while the remaining 28 delve into various privacy attacks. Additionally, we found 58 papers dedicated to exploring privacy protection strategies for LLMs. We classified them according to different phases: 11 during pre-training, 23 during fine tuning, and 24 in inference phase. An analysis of publication trends shows that the majority of these papers, representing 58.57\%, were published in 2023, with only 30 released in between 2021 and 2022, indicating a significant recent interest in the topic. Notably, there are also 5 cutting-edge studies from 2024, which underscores the ongoing and dynamic nature in this crucial area of research. 

\vspace{2mm}
\noindent\textbf{Structuring and Analysis:} Fig. \ref{fig:privacy-threats} presents the organizational structure of this study, which outlines the current privacy threats faced by LLMs and its corresponding protections as well as the relevance between privacy threats and defense technologies. In the section on privacy threats, this paper reviews existing research from two dimensions: privacy attacks and privacy leakage, detailing common attack methods and instances of privacy leakage in LLMs. Regarding privacy protection approaches, we systematically summarize them according to the three stages of LLMs: pre-training, fine-tuning, and inference. And we summarize the key privacy protection technologies, including data sanitization, federated learning, differential privacy, homomorphic encryption, and secure multi-party computation. Finally, we establish a connection between these key technologies and the privacy threats they may defend against, providing a framework for understanding the data privacy in LLMs.

\subsection{Overview}

Figure \ref{fig:privacy-threats} offers an intricate portrayal of privacy concerns, encapsulating both privacy leaks and attacks, alongside the tailored defensive technologies deployed at various phases of LLMs lifecycle, including pre-training, fine-tuning, and inference stages. 
\begin{itemize}
    \item \textbf{Privacy Threats (\S\ref{sec:threats})}: We first conduct a literature review on privacy threats against LLMs. Based on whether the attackers are active or passive, we further categorize the threats into two groups: privacy leakage, where the attackers passively collect sensitive information due to vulnerabilities, and privacy attacks, where the attackers actively break LLMs to access sensitive information.
    \item 
\textbf{Privacy Protections (\S\ref{subsec:defense1} \& \S\ref{subsec:defense2}):} Based on where privacy protection is located, we can group them into three categories: privacy protection in Pre-Training (\S\ref{subsec:defense1.1}), privacy protection in fine-tuning (\S\ref{subsec:defense1.2}), and privacy inferences (\S\ref{subsec:defense2}). Among them, privacy protection in inferences can be further grouped based on the methods adopted (e.g., whether it is a cryptography-based approach). In each of these protections, we first introduce the techniques at a high level; then, we explain how they can be used in LLMs (see those \textbf{Tech Tips}), and finally, we provide a detailed literature review.
\end{itemize}

% \add{The illustration methodically segments the privacy threats targeted by this research on the left, while delineating the corresponding countermeasures on the right, engineered to neutralize such vulnerabilities. Subsequent sections of this paper will meticulously examine the implementation and efficacy of these privacy safeguarding tactics within LLM frameworks, offering an in-depth guide for safeguarding privacy in LLM applications.}
% \ZY{Discuss Figure 4 at a very high level. Also, use this figure to guide the following sections.}

% This paper aims to answer several key questions to provide a thorough understanding of privacy implications of Large Language Models (LLMs).
% \begin{itemize}
%     \item  \textbf{Privacy Threats:}  What specific privacy threats arise from the use of LLMs? 
%     \item  \textbf{Privacy-Preserving Techniques:} How do privacy protection techniques manifest at different stages of the LLM lifecycle (e.g., pre-training, fine-tuning, inference)? How effective are these techniques in mitigating privacy risks? 
%     \item  \textbf{Future Directions:} What are the limitations and challenges associated with implementing privacy-preserving measures in LLMs? What are the emerging trends and developments in LLM privacy research?
% \end{itemize}

 \section{Privacy Leakage and Privacy Attacks in LLMs}
 \label{sec:threats}

We undertake a literature review focusing on privacy threats against LLMs. We categorize these threats into two groups based on the attackers' activity: privacy leakage, wherein attackers passively collect sensitive information due to vulnerabilities, and privacy attacks, wherein attackers actively breach LLMs to access sensitive information
 
% % Pre-training def.
% In this section, we discuss data privacy threats in large language models. We primarily focus on privacy leakage and privacy attacks. Privacy leakage focuses on the risk of inadvertently leaking sensitive information, which may stem from the extensive datasets used during the model's training process, or the model's memory capability that leads to retaining information that should not be disclosed. Privacy attacks explore how adversary might exploit LLMs to launch attacks aimed at obtaining or inferring sensitive information from the training data.

% \begin{itemize}
%     \item  \subsubsection{Data Privacy:} During pre-training, large language models are typically trained on vast amounts of text data. If this data contains sensitive or personally identifiable information (PII) without proper anonymization or privacy protection measures, there's a risk of privacy leakage.
% \end{itemize}

\subsection{Privacy Leakage (Passive)} 
\label{subsec:passive}
\subsubsection{\textbf{Sensitive Query}} Users may input queries containing sensitive or personally identifiable information (PII) into LLMs. For example, asking questions about medical conditions, financial situations, or personal relationships could reveal private details about the user's life.
% Privacy of LLM
If users input sensitive information as prompts, there arise concerns regarding data privacy \cite{kshetri2023cybercrime, zamfirescu2023johnny}. For example, Samsung Electronics staff provided sensitive corporate data when interacting with ChatGPT. 
% Privacy of LLM Plugins
Besides, various LLM plugins also raise privacy concerns of user's sensitive data. Iqbal \textit{et al.}~\cite{iqbal2023llm} proposed a systematic framework to evaluate the security, privacy, and safety of third-party plugins integrated into LLM platforms, focusing on OpenAI's ChatGPT ecosystem. Some plugins were found to collect excessive user data, including personal and sensitive information. Some plugins did not provide clear details on how they use user data, potentially violating privacy policies.

\subsubsection{\textbf{Contextual Leakage}} Even seemingly innocuous queries could indirectly reveal sensitive information about the user when combined with other contextual factors. For instance, asking about nearby landmarks or local events could inadvertently disclose the user's location or activities. Over time, repeated interactions with the model could lead to the accumulation of enough information to uniquely identify the user, posing a risk to privacy.
% Personal attributes
The study \cite{staab2023beyond} focuses on the capabilities of LLMs to infer personal attributes from text, particularly in the context of privacy concerns and the threat of privacy-invasive chatbots. They evaluated LLMs' ability to infer personal attributes (like location, occupation, age, gender, etc.) from text on the PersonalReddit dataset, containing 520 profiles with 5814 comments. They evaluated 9 state-of-the-art LLMs on the PR dataset, with GPT-4 achieving top-1 accuracy of 84.6\% and top-3 accuracy of 95.1\%.

\subsubsection{\textbf{Personal Preferences Leakage}}  LLMs may infer personal preferences, interests, or characteristics of users based on their queries and interactions. This could result in targeted advertisements, personalized recommendations, or other tailored content that may reveal private aspects of the user's life. 
% LLM recommendation
For example, LLMs represent a significant asset to recommender systems, offering advantages in delivering personalized recommendations \cite{lyu2023llm}. Besides, these models have the potential to refine or establish new methodologies for sequential recommendation \cite{harte2023leveraging}, which could inadvertently reveal users' personal preferences, thereby raising privacy concerns.

 % \subsubsection{Adversarial Inputs} Adversaries may craft inputs or queries designed to exploit vulnerabilities in LLMs and elicit unintended or malicious responses. These adversarial inputs could lead to privacy breaches by revealing sensitive information or causing the model to generate biased or inappropriate outputs. Adversaries may try to infer sensitive information about individuals or organizations by querying LLMs during the inference phase. For example, by analyzing the model's responses to queries, adversaries could infer personal preferences, opinions, or intentions of individuals interacting with the model, leading to privacy breaches.

During the utilization of LLMs, individuals may unintentionally disclose their privacy, whether through direct or indirect means. Beyond the direct provision of sensitive information, providers of services can extrapolate intricate user attributes and preferences, thereby gaining access to sensitive data via data analysis methods.

\subsection{Privacy Attacks (Active)}
\label{subsec:active}
 
 \subsubsection{\textbf{Backdoor Attacks (Data Poisoning Attacks) on Pre-Training}} During the pre-training phase, the adversary manipulates the training data, introducing poison into the dataset. Subsequently, this tainted training data is disseminated on the internet, where unwitting developers procure and employ it for training their models. Consequently, the models become infused with covert backdoors, thereby compromising their integrity and security.  Adversaries can exploit backdoors to exfiltrate sensitive or private information processed by the LLMs \cite{li2021backdoor}. This could include personal data, confidential documents, or proprietary information, leading to privacy breaches and potential violations of data protection regulations. Backdoors allow adversaries to manipulate the output of LLMs, potentially leading to the generation of misleading or harmful content. This can have detrimental effects on users' privacy, particularly if the manipulated content contains false information or malicious intent. Yang \textit{et al} \cite{yang2021careful} shed light on a critical security vulnerability in NLP models, introducing a data-free backdoor attack that could subvert the integrity of word embeddings by altering a single embedding vector. POISONPROMPT \cite{yao2023poisonprompt} emerges as a novel backdoor attack strategy, demonstrating to its capability to compromise both hard and soft prompt-based LLMs. Furthermore, Huang \textit{et al} \cite{huang2023composite} introduced a stealthy Composite Backdoor Attack (CBA) that scatters multiple trigger keys across different prompt components. CBA ensures activation only when all triggers are present, demonstrating high effectiveness in NLP and multimodal tasks while maintaining model accuracy.

 \subsubsection{\textbf{Backdoor Attacks (Data Poisoning Attacks) on Fine Tuning}} Adversaries may inject poisoned or adversarial examples into the fine-tuning dataset to manipulate the behavior of LLMs. These poisoned examples could introduce biases or vulnerabilities into the model, leading to compromised performance or biased outputs that violate privacy and fairness principles. Research by Wan \textit{et al} \cite{wan2023poisoning} has revealed that instruction-tuned LMs, such as ChatGPT, are vulnerable to backdoor attacks where adversaries can manipulate model behavior by tainting training datasets with malicious examples. These poisoned models then exhibit erroneous behavior when exposed to specific trigger phrases, leading them to produce a predetermined target label in classification tasks. In a similar vein, Xu \textit{et al} \cite{xu2023instructions} demonstrated that attackers can subvert model behavior by interspersing legitimate data with malicious instructions, achieving high success rates of exploitation across various NLP datasets. Furthermore, these attacks can be engineered to elicit targeted or even harmful responses on specific topics. For example, Yan \textit{et al} \cite{yan2023backdooring} have shown that it is possible for adversaries to implant Virtual Prompt Injection (VPI) backdoors into models through tainted instruction tuning data, granting them the capability to finely control the model's outputs in response to carefully chosen triggers.

 \subsubsection{\textbf{Membership Inference Attacks on Pre-Training}} In a membership inference attack~\cite{shokri2017membership,huang2021damia}, an adversary attempts to determine whether a specific individual's data was included in the training dataset used to train an LLM. By analyzing the model's outputs or responses to queries, the attacker can infer whether certain data samples were part of the training data. This can lead to privacy breaches if sensitive information about individuals is inferred from the model's behavior. A study by Mireshghallah \textit{et al} \cite{mireshghallah2022quantifying} has highlighted the high susceptibility of Masked Language Models (MLMs) to privacy attacks, demonstrating this through likelihood ratio membership inference attacks that utilize an additional reference MLM. However, considering the unrealistic assumption of reference-based models, Mattern \textit{et al} \cite{mattern2023membership} proposed an alternative method known as neighbourhood attacks, which compare scores with synthetic texts. In another development, Shi \textit{et al} \cite{shi2023detecting} introduced WIKIMIA benchmark and MIN-K PROB method, which they claimed improved detection by 7.4\% over previous methods. Despite these advancements, Duan \textit{et al} \cite{duan2024membership} evaluated membership inference attacks on the pre-training data of LLMs trained on the Pile and found that the success rates of the aforementioned attack methods were limited due to the combination of large datasets and few training iterations, as well as a fuzzy boundary between members and non-members.

 \subsubsection{\textbf{Membership Inference Attacks on Fine Tuning}} Membership Inference Attacks aim to reveal whether specific data samples have been incorporated into the training set of the model. In the context of LLM fine-tuning, adversaries may discern patterns that suggest whether those inputs were part of the training data by meticulously analyzing the model's responses to certain inputs. Accurate execution of such an inference by attackers could lead to the compromise of the model's training data confidentiality. Mireshghallah \textit{et al.} \cite{mireshghallah2022empirical} conducted an empirical investigation that examined significant variation in vulnerability to membership inference of different fine-tuning methods for LLMs. Their findings indicated fine-tuning model heads proves most susceptible, while using smaller adapters shows reduced attack susceptibility. 
 Moreover, Jagannatha \textit{et al.} \cite{jagannatha2021membership} focused on fine-tuned clinical language models (CLMs) and their exposure to MIAs. They demonstrated that the scale of the model plays a crucial role in its privacy risks, with smaller models generally exhibiting lower vulnerability compared to larger architectures. Building on these insights, Fu \textit{et al.} \cite{fu2023practical} introduced a novel approach to MIA in fine-tuned LLMs. Their proposed method, Self-calibrated Probabilistic Variation (SPV-MIA), leverages memorization rather than overfitting as a reliable indicator of membership. Additionally, they presented a self-prompt strategy for constructing a comparable dataset for the reference model, aiming to enhance the practicality and effectiveness of MIAs against fine-tuned LLMs.
 
 \subsubsection{\textbf{Model Inversion (Data Reconstruction) Attacks}} In a model inversion attack, an adversary attempts to reconstruct or reverse-engineer the training data used to train an LLM based on its outputs or internal representations. By analyzing the model's parameters, gradients, or generated text, the attacker aims to recover sensitive information contained in the training data, such as personal communications, financial records, or proprietary documents. Song \textit{et al} \cite{song2020information} demonstrated this through the development of such an attack. And the study by Carlini \textit{et al} \cite{carlini2021extracting} on GPT-2 demonstrates that adversaries can extract individual training examples through training data extraction attacks. Following this, Lehman \textit{et al} \cite{lehman2021does} further investigated the risk of model inversion attacks on a BERT model trained on sensitive EHR data. Surprisingly, they found that simple probing methods failed to extract sensitive information, indicating a potential safety margin for releasing such model weights. However, \emph{Text Revealer} \cite{zhang2022text} was designed by Zhang \textit{et al}, which is the first model inversion attack specifically designed for reconstructing private texts from transformer-based text classification models. Its attack leverages external datasets and GPT-2 to generate fluent, domain-specific text, optimizing perturbations to the hidden state based on feedback from the target model.

 \subsubsection{\textbf{Attribute Inference Attacks}} Attribute inference attacks involve inferring sensitive attributes or characteristics of individuals from fine-tuned LLMs. For example, an attacker may attempt to infer demographic information, such as age, gender, or ethnicity, based on the language patterns or topics discussed in the model's generated text \cite{li2023privacy1}. This can lead to privacy violations and discrimination against individuals based on inferred attributes. In a comprehensive study, Pan \textit{et al.} \cite{pan2020privacy} systematically examined the privacy risks associated with 8 state-of-the-art language models. Their examination is anchored on 4 diverse case studies that focus on the threat of attribute inference attacks. The findings are compelling: these state-of-the-art models are indeed susceptible to revealing sensitive details, which include personal identifiers such as identity, genetic information, health data, and geographical locations. This vulnerability stems from the potential for adversaries to reverse-engineer the embeddings within these models. Building on this concern, Staab \textit{et al.} \cite{staab2023beyond} employed Reddit profiles to showcase that LLMs can accurately infer a variety of personal attributes. Remarkably, these models surpass human performance in terms of both efficiency and speed, underscoring the urgent need for effective privacy safeguards in model development.

 \subsubsection{\textbf{Model Stealing Attacks}} Adversaries may attempt to steal or replicate fine-tuned models trained on proprietary or sensitive datasets. By querying the model and observing its responses, adversaries can extract information about the model's parameters or internal representations, enabling them to reconstruct or replicate the model without access to the original training data. Krishna \textit{et al.} \cite{krishna2019thieves} demonstrated the feasibility of model stealing attacks in NLP, showing that adversaries can reconstruct victim models using only random word sequences and task-specific heuristics, without requiring real training data. This exploit is enabled by the widespread use of transfer learning methods in NLP. And then, Truong \textit{et al.} \cite{truong2021data} advanced the field with their proposal of data-free model stealing techniques. These methods overcome the need for surrogate datasets, enabling accurate replication of valuable models with limited queries. Besides, Sha \textit{et al.} \cite{sha2024prompt} introduced a novel prompt stealing attack against LLMs, leveraging generated answers to reconstruct well-designed prompts. It involves a two-pronged approach: a parameter extractor dissects prompt types and characteristics, while a prompt reconstructor generates reverse-engineered prompts with notable efficacy.

\section{Privacy Protection in Pre-Training and Fine-Tuning}
\label{subsec:defense1}

Privacy protection in pre-training and fine-tuning of LLMs is paramount in safeguarding sensitive data while ensuring model effectiveness. Incorporating techniques such as differential privacy, data cleaning, and federated learning can mitigate privacy risks.

\subsection{Privacy Protection in Pre-Training}
\label{subsec:defense1.1}

\subsubsection{\textbf{Data Cleaning}}
Data cleaning enhances data quality by rectifying errors and inconsistencies, serving as a foundational step that also plays a critical role in privacy protection by implementing anonymization, data minimization, and secure practices to safeguard sensitive information. To be more specific, we can remove or generalize personally identifiable information (PII) such as names, addresses, social security numbers, etc., to make it harder to identify individuals in the dataset (e.g., we can also mask sensitive information by replacing it with non-sensitive placeholders or pseudonyms while still preserving the structure and relationships within the dataset); we can aggregate data at a higher level to reduce the risk of re-identification. For example, instead of storing individual inference query details, and aggregate queries by day or week.

\begin{mybox}[boxsep=0pt,
	boxrule=1pt,
	left=4pt,
	right=4pt,
 	top=4pt,
 	bottom=4pt,
	]
 
\textbf{Tech Tips:} When utilizing data cleaning techniques for privacy protection in LLMs, it's essential to prioritize thorough data sanitization before fine-tuning the model for specific tasks. Anonymizing or pseudonymizing sensitive information, and aggregating data to reduce granularity are key strategies to safeguard individual privacy.  
\end{mybox}
OpenAI\cite{brown2020language} underscores the thorough measures implemented to elevate the quality and security of their training data. They utilize filtering and fuzzy deduplication techniques to remove personally identifiable information from the corpora utilized for model training. This methodology not only purifies the data but also secures a heightened level of privacy protection. These measures are also employed in\cite{ouyang2022training}. Anthropic\cite{bai2022training} adopts a strategic approach in their training methodologies, focusing on the exclusive use of beneficial human feedback data to develop AI assistants. This selective data utilization guarantees the creation of assistants that are intrinsically helpful and non-harmful, founded on a foundation of entirely positive interactions. Additionally, their commitment to fostering AI behavior that aligns with constitutional and ethical standards is further highlighted in\cite{bai2022constitutional}. Kandpal \textit{et al.} \cite{kandpal2022deduplicating} demonstrated that removing duplicated sequences from training data significantly reduces the vulnerability of language models to privacy attacks, such as those allowing adversaries to recover memorized information\cite{carlini2021extracting}. Through empirical analysis, the authors show that duplication in training data is a key factor contributing to these privacy risks. By deduplicating the training sets, the models become less likely to regenerate sensitive or specific information, hence improving their security against such attacks without compromising the model's performance.

\subsubsection{\textbf{Federated Learning}}

Federated learning revolutionizes machine learning by decentralizing the training process, enabling model training across multiple edge devices or servers while preserving data privacy. Initially, a global model is distributed to participating devices, which independently train the model using their local data. Instead of sending raw data to a central server, only model updates are transmitted, ensuring user privacy as data remains localized. These updates are aggregated at the central server to refine the global model iteratively, leading to continuous improvement without compromising privacy. Federated learning thus offers a paradigm shift, promoting collaborative machine learning in privacy-sensitive environments by leveraging distributed data processing and maintaining data locality.

\begin{mybox}[boxsep=0pt,
	boxrule=1pt,
	left=4pt,
	right=4pt,
 	top=4pt,
 	bottom=4pt,
	] 
 
\textbf{Tech Tips:} In the pre-training of LLMs, federated learning offers a privacy-centric approach by eliminating the need for centralized data storage. Training occurs on local devices, with only model parameters or updates sent to a central server for aggregation. This method keeps personal data on its original device, drastically reducing data breach risks and addressing privacy and security concerns associated with centralized storage. \looseness=-1
\end{mybox}

Chen \textit{et al.}\cite{chen2023federated} introduced a federated learning framework for LLMs that focuses on privacy without sacrificing performance, incorporating federated pre-training to securely utilize decentralized data for improved privacy, security, and model generalization. 
Yu \textit{et al.} \cite{yu2023federated} developed Federated Foundation Models to enhance privacy in collaborative learning, focusing on the entire lifecycle of foundation models with federated learning. They tackle privacy, performance, and scalability, paving the way for future research on privacy-preserving, personalized models.

 \begin{tcolorbox}[colback=blue!2!white,colframe=blue!20!black,title=Finding: Federated learning is not enough]
    Federated learning protects data privacy across various participants by decentralizing the training process, where data remains on users' devices and only model updates are shared. However, it's not entirely secure against privacy breaches; malicious servers could potentially extract private user data from shared gradients. To bolster security, federated learning often integrates additional privacy-preserving techniques such as differential privacy, secure multi-party computation, homomorphic encryption, and adversarial training. These methods collectively enhance the robustness of privacy protection in federated learning frameworks.
\end{tcolorbox}

\subsubsection{\textbf{Differential Privacy}}
Differential privacy is a technique for protecting data privacy, particularly in the fields of statistical release and data analysis. Its purpose is to allow researchers to extract useful statistical information from an entire dataset without revealing any individual data. Differential privacy achieves this by adding a certain amount of random noise to the data, ensuring that even if attackers have complete background knowledge except for the target dataset, they cannot determine whether the dataset contains information about a specific individual. We can define differential privacy as follows:
\begin{definition}
Given two datasets $D_1$ and $D_2$, that differ by only one element (i.e., they are ``adjacent datasets''), a randomized algorithm $A$ satisfies $\epsilon$-differential privacy if and only if for all output sets $S$ from the algorithms on $D_1$ and $D_2$, the following holds:
\begin{equation}
\frac{\operatorname{Pr}\left[\mathcal{A}\left(D_1\right) \in S\right]}{\operatorname{Pr}\left[\mathcal{A}\left(D_2\right) \in S\right]} \leq e^\epsilon
\end{equation}
where $\operatorname{Pr}\left[\mathcal{A}\left(D_1\right) \in S\right]$ represents the probability that the result of running algorithm $\mathcal{A}$ on dataset $D_1$ falls within the set $S$. $\epsilon$ is a non-negative parameter known as the privacy budget. The smaller the $\epsilon$, the higher the level of privacy protection, but this may reduce the utility of the data. $e$ is the base of the natural logarithm, approximately equal to $2.71828$.

Since the algorithm $\mathcal{A}$ is random, differential privacy can ensure that for adjacent datasets (i.e., datasets that differ by only one element), the output of an algorithm is ``almost identical.'' This means that it is nearly impossible to infer any specific information about an individual from the output. By adjusting the value of $\epsilon$, a trade off can be realized between data privacy protection and data utility.
\end{definition}

\begin{mybox}[boxsep=0pt,
	boxrule=1pt,
	left=4pt,
	right=4pt,
 	top=4pt,
 	bottom=4pt,
	]
 
\textbf{Tech Tips:} Integrating differential privacy into the pre-training process of LLMs involves adding noise to the training data or model updates to safeguard individual privacy while maintaining effective model training. This can be achieved by injecting random noise into training data or perturbing gradients during backpropagation. Adaptive noise mechanisms dynamically adjust noise levels based on data sensitivity and privacy budgets. Careful management of the privacy budget ensures desired privacy levels are maintained. 
\end{mybox}

Hoory \textit{et al.} \cite{hoory2021learning} examined the application of differential privacy to pre-trained language models. It focuses on evaluating and enhancing the performance of these models under privacy constraints. 
Du \textit{et al.} \cite{du2021dp} focused on providing differential privacy in forward propagation for large-scale models. It addresses the challenge of protecting data privacy while performing forward propagation in large models. 
Li \textit{et al.} \cite{li2021large} argued that LLMs can be effective learners under differential privacy constraints. It explores techniques to optimize model performance while adhering to privacy standards.

\subsection{Privacy Protection in Fine Tuning}
 \label{subsec:defense1.2}
 \subsubsection{\textbf{Federated Learning}} 
 Federated learning transcends its initial application in pre-training, proving equally effective in the fine-tuning phase. This expanded application not only extends its utility but also underscores its versatility in bolstering privacy protection. By operating across data, models, and commands, federated learning presents a holistic solution, showcasing its comprehensive applicability and potential for addressing privacy concerns in diverse contexts.
 
\begin{mybox}[boxsep=0pt,
	boxrule=1pt,
	left=4pt,
	right=4pt,
 	top=4pt,
 	bottom=4pt,
	] 
\textbf{Tech Tips:} Similarly, in the fine-tuning phase, federated learning is employed by initially distributing the pre-trained global model to edge devices or local servers where fine-tuning tasks are performed. On each device or server, the global model is fine-tuned using locally held data pertinent to the specific task.  
\end{mybox}
Xu \textit{et al.}\cite{xu2024fwdllm} and Zhang \textit{et al.}\cite{zhang2024building} integrated federated learning into the fine-tuning of LLMs to significantly enhance privacy protection. Their approaches focus on keeping sensitive data on the user's device, eliminating the need for direct data transmission and sharing. By employing advanced privacy-preserving techniques such as differential privacy, secure aggregation, and homomorphic encryption, they ensure that user privacy is safeguarded during the fine-tuning process. Sun \textit{et al.}\cite{sun2023fedbpt} introduced FedBPT, a federated learning framework for privacy-preserving prompt tuning in language models, optimizing prompts locally and sharing only updates to minimize communication overhead and ensure data privacy. This method facilitates secure, collaborative model enhancement without exposing sensitive data. Zhao \textit{et al.} \cite{zhao2023privacy} enhanced privacy in model fine tuning across decentralized nodes by aggregating local updates into a central model without centralizing data, effectively keeping sensitive information local and mitigating data breach risks while leveraging collaborative learning benefits. Fan \textit{et al.}\cite{fan2023fatellm} presented an approach that combines federated learning with knowledge distillation and parameter-efficient fine-tuning in LLMs to ensure privacy. They also introduce secure aggregation for safely merging model updates, enabling collaborative, privacy-preserving learning across different organizations. 

 \begin{tcolorbox}[colback=blue!2!white,colframe=blue!20!black,title=Finding: Federated Learning in Pre-Training V.S. in Fine-Tuning]
 {In federated learning, pre-training employs extensive, general datasets for foundational language comprehension through distributed learning, emphasizing data privacy. Fine-tuning, however, focuses on specialized tasks using targeted datasets, prioritizing personalized optimization and stricter privacy on local devices. The technical needs for privacy protection distinctly vary between these stages. However, most research on addressing privacy issues in LLMs through federated learning focuses on optimizing the computational and communication overhead. These studies either claim applicability to both pre-training and fine-tuning phases or claim relevance to a specific phase without making targeted adjustments or designs for that stage. This highlights a gap: the need for precise, stage-specific optimization and design in federated learning for LLMs, essential for improving privacy protection's effectiveness and efficiency at different stages.
 }
\end{tcolorbox}
 
 \subsubsection{\textbf{Differential Privacy}} 
The approaches primarily employ differential privacy techniques to handle privacy-sensitive tuning data, thereby enabling secure and private inference. These approaches focus on balancing the data utility in model tuning with the data privacy \cite{hoory2021learning,du2021dp,li2021large,behnia2022ew,shi2022just,wu2022adaptive,majmudar2022differentially,du2023dp,li2023privacy1,mai2023split}. 
Behnia \textit{et al.}\cite{behnia2022ew} introduced EW-Tune, a framework for fine-tuning LLMs with differential privacy guarantees. EW-Tune employed the Edgeworth accountant method, offering finite-sample privacy guarantees suitable for the fine-tuning context. It solves the problem of how to fine-tune LLMs on private data without compromising privacy. 
Shi \textit{et al.}\cite{shi2022just} presented a framework for enhancing the privacy of LLMs without significantly compromising their utility. The proposed approach, Just Fine-tune Twice (JFT), focuses on selectively applying differential privacy (SDP) to only the sensitive parts of data, based on a policy function. This is achieved through a two-phase fine-tuning process: first with redacted data and then with original data using a privacy-preserving mechanism. This method is shown to be effective for transformer-based models and addresses limitations of prior SDP applications.
Wu \textit{et al.} \cite{wu2022adaptive} designed an Adaptive Differential Privacy (ADP) framework for language model training. It estimates the privacy probability of linguistic items without resorting to the prior privacy information and designs a novel Adam algorithm to adaptively adjust the degree of differential privacy noise, potentially improving model utility while maintaining privacy.  
Li \textit{et al.} \cite{li2023privacy1} explored a method for prompt tuning LLMs in a privacy-preserving manner. This approach seeks to leverage the power of large models while safeguarding user privacy.

% \subsection{Learning-based Approaches}
 
 \subsubsection{\textbf{Knowledge Unlearning}} 
  Knowledge unlearning, also known as machine unlearning, is a strategy aimed at bolstering privacy within machine learning models, especially LLMs\cite{liu2024rethinking}.
  When a machine learning model is trained on data, it learns patterns and correlations present in that data. However, sometimes these patterns may inadvertently encode sensitive information about individuals. If the model retains this information, it can pose privacy risks when the model is deployed in real-world applications, especially in scenarios where the model may be exposed to sensitive data. Knowledge unlearning techniques aim to mitigate these risks by selectively forgetting or removing sensitive information from the model. \looseness=-1

  \begin{mybox}[boxsep=0pt,
	boxrule=1pt,
	left=4pt,
	right=4pt,
 	top=4pt,
 	bottom=4pt,
	] 
 
\textbf{Tech Tips:}
  In the fine-tuning stage, it functions by ensuring that the model does not hold onto or disclose sensitive details learned during its initial training phases. This process involves retraining the model to eliminate its memory of certain information, effectively reducing the risk of privacy breaches while maintaining or enhancing the model's performance.  
 \end{mybox}
 
 Zhang \textit{et al.}\cite{zhang2023right} analyzed the Right to be Forgotten in LLMs, identifying the unique legal and technological hurdles and proposing solutions like differential privacy and machine unlearning to balance privacy with technological progress. Chen \textit{et al.}\cite{chen2023unlearn} introduced an efficient unlearning technique for LLMs using unlearning layers within transformers, enabling precise data removal without retraining and effectively managing sequential deletion requests with minimal performance loss. Jang \textit{et al.}\cite{jang2022knowledge} proposed a targeted unlearning method for LMs through gradient ascent on specific sequences, offering an efficient way to erase sensitive information while preserving overall model performance. Eldan \textit{et al.}\cite{eldan2023whos} detailed a novel unlearning approach for LLMs by fine-tuning on datasets modified to omit targeted knowledge, employing reinforcement bootstrapping to forget information without compromising model integrity.

\subsubsection{\textbf{Offsite Tuning}}
Offsite tuning, detailed by Xiao \textit{et al.}\cite{xiao2023offsitetuning}, refines the adaptability of models to specific tasks, prioritizing data privacy through the deployment of lightweight adapters and compressed emulators for localized adjustments.

\begin{mybox}[boxsep=0pt,
	boxrule=1pt,
	left=4pt,
	right=4pt,
 	top=4pt,
 	bottom=4pt,
	]

 \textbf{Tech Tips:}
This innovative method transmits only essential components to the data owner for offsite tuning, thereby avoiding the exposure of the entire model and ensuring that sensitive data remains under the data owner's control. This significantly lowers the risk of privacy breaches. The adapter, fine-tuned with local data, is updated without direct data exposure and seamlessly reintegrated into the foundation model, effectively safeguarding data privacy throughout the adaptation process.
 \end{mybox}

\section{Privacy Protection in Inference}
\label{subsec:defense2}
During the inference process of LLMs, the issue of privacy leakage has garnered widespread attention. To address this issue, researchers have developed numerous strategies to ensure privacy security during the inference phase. In this section, we summarize the privacy protection approaches for the inference stage of LLMs, focusing on various approaches including encryption-based privacy protection approaches, privacy protection approaches through detection, and hardware-based approaches.

% \begin{itemize}
%     \item  \subsubsection{Contextual Generation:} When using the LLM for inference, such as generating responses in a chatbot or completing text prompts, there's a risk that the model may produce outputs containing sensitive information learned during pre-training or fine-tuning. For example, if the model has been exposed to sensitive data during training, it may generate responses that inadvertently disclose this information.
%     \item  \subsubsection{Adversarial Attacks:} Adversarial attacks can be launched against LLMs during inference, aiming to elicit sensitive information or manipulate model behavior in unintended ways. These attacks can exploit vulnerabilities in the model's architecture or training process to extract or infer private details.
% \end{itemize}

\begin{table*}
\caption{Private Inference Approaches} 
\label{table: performance comparison}
\renewcommand\arraystretch{1.3}
\begin{center}
\begin{threeparttable}
    \tabcolsep=0.2cm
    \footnotesize
    \begin{tabular}{lcccccc}
    \toprule[1pt]
    Schemes         & Tools & Improved Components      & \begin{tabular}[c]{@{}c@{}}Matrix\\  Multiplication\end{tabular} & \begin{tabular}[c]{@{}c@{}}Nonlinear\\ to Polynomial\end{tabular}
    & Threat Model & Experiments on      \\ 
    \midrule[0.5pt]
    THE-X$\dag$ \cite{chen2022x}                             & HE               & \begin{tabular}[c]{@{}c@{}} GELU, SoftMax, \\ LayerNorm\end{tabular}    & $\CIRCLE$             & $\Circle$            & CPA                 & Bert-tiny \\ 
    Iron \cite{hao2022iron}                                  & HE               & \begin{tabular}[c]{@{}c@{}} GELU, SoftMax, \\ LayerNorm\end{tabular}    & $\CIRCLE$             & $\Circle$            & Honest-but-curious  & Bert \\ 
    Bumblebee \cite{lu2023bumblebee}                         & HE               & SoftMax, LayerNorm                                                      & $\CIRCLE$             & $\Circle$            & Static semi-honest  & \begin{tabular}[c]{@{}c@{}}Bert-base/Large, GPT2-base, \\ LLaMA-7B, ViT-base\end{tabular} \\ 
    Zimerman \textit{et al.}$\dag$ \cite{zimerman2023converting}      & HE               & GELU, Softmax                                                           & $\Circle$             & $\CIRCLE$            & CPA                 & BerT-like \\ 
    Liu \textit{et al.} \cite{liu2023llms}                            & HE, MPC          & \begin{tabular}[c]{@{}c@{}} GELU, SoftMax, \\ LayerNorm\end{tabular}    & $\CIRCLE$             & $\Circle$            & Semi-honest         & \begin{tabular}[c]{@{}c@{}}BERT-Tiny, BERT-Medium,  \\ RoBERTa-Base\end{tabular} \\ 
    Wang \textit{et al.} \cite{wang2022characterization}              & MPC              & \begin{tabular}[c]{@{}c@{}} SoftMax, Embedded \\ Tables\end{tabular}    & $\Circle$             & $\Circle$            & Semi-honest         & XLM, ViT \\ 
    CipherGPT \cite{hou2023ciphergpt}                        & MPC              & GELU                                                                    & $\CIRCLE$             & $\Circle$            & Semi-honest         & GPT2 \\ 
    East  \cite{ding2023east}                                & MPC              & SoftMax, LayerNorm                                                      & $\Circle$             & $\CIRCLE$            & Semi-honest         & BerT-like \\ 
    Privformer \cite{akimoto2023privformer}                  & MPC              & Sigmoid                                                                 & $\CIRCLE$             & $\Circle$            & Honest majority     & Transformer \\ 
    Puma \cite{dong2023puma}                                 & MPC              & GELU, SoftMax                                                           & $\Circle$             & $\CIRCLE$            & Semi-honest         & \begin{tabular}[c]{@{}c@{}}Bert-Base/Large,  \\ GPT2-Base/Medium/Large, \\ Roberta-Base, LLaMA-7B \end{tabular} \\ 
    Sigma \cite{gupta2023sigma}                              & FSS              & GELU, SoftMax                                                           & $\Circle$             & $\CIRCLE$            & Semi-honest static  & \begin{tabular}[c]{@{}c@{}} Bert-Tiny/Base/Large, \\ GPT2, GPT2-Neo \end{tabular} \\ 
    Majmuda \textit{et al.} \cite{majmudar2022differentially}         & DP               & SoftMax                                                        & $\Circle$             & $\Circle$            & Semi-honest           & RoBERTa-style \\ 
    Dp-forward  \cite{du2023dp}                              & DP               & Embedding   & $\Circle$             & $\Circle$            &  Semi-honest  & Bert \\ 
    Mai \textit{et al.} \cite{mai2023split}                           & DP               & Embedding                                                                   & $\Circle$             & $\Circle$            & \begin{tabular}[c]{@{}c@{}}   Attribute \\ inference attack\end{tabular} & Bert, GPT2, T5 \\ 
     Textfusion \cite{zhou2022textfusion}                    & \begin{tabular}[c]{@{}c@{}} Token fusion \end{tabular}      & Tokenizer            & $\Circle$             & $\Circle$            & \begin{tabular}[c]{@{}c@{}} Text reconstruction \\  attack \end{tabular}   &  Bert-Base, Bert-Large \\ 
      Yuan \textit{et al.} \cite{yuan2023secure}                      & Permutation     & \begin{tabular}[c]{@{}c@{}} RELU, SoftMax, \\ LayerNorm\end{tabular}     & $\CIRCLE$             & $\Circle$            & -                    & Transformer, LLaMa\\ 
    \bottomrule[1pt]
    \end{tabular}
    	\begin{tablenotes}
		    \item[CPA] Chosen plaintext attacks. 
            \item[$\dag$] Note that the CKKS homomorphic encryption scheme might be vulnerable to passive attacks. \cite{li2021security} 
		\end{tablenotes}
\end{threeparttable}
\end{center}
\end{table*}

\subsection{Cryptography-based Approaches}

 \subsubsection{\textbf{Homomorphic Encryption}} 
Homomorphic encryption \cite{acar2018survey} is a cryptographic technique that allows for computations to be performed on ciphertexts, ensuring that the result, when decrypted, is identical to the result of the same operations performed on the plaintext. This encryption method is key in enabling data to be processed while maintaining its encrypted state, adding a new dimension to data privacy and security. Homomorphic encryption is primarily categorized into three types:
\begin{itemize}
\item Partial Homomorphic Encryption (PHE): Supports one type of operation (usually addition or multiplication) on ciphertexts.
\item Somewhat Homomorphic Encryption (SWHE): Allows a limited number of operations on ciphertexts.
\item Fully Homomorphic Encryption (FHE): The most powerful, supporting an unlimited number of both addition and multiplication operations on ciphertexts.
\end{itemize}

To better understand homomorphic encryption algorithms, we provide the following definition.
\begin{definition}
An encryption scheme is considered homomorphic over an operation $\circ$ if it satisfies a specific mathematical property. Specifically, it supports the following equation:
\begin{equation}
    E(m_1) \circ E(m_2)=E(m_1 \circ m_2), \ \  \forall m_1,m_2 \in \mathcal{M}
\end{equation}
\end{definition}
Here, $E$ represents the encryption algorithm, $\mathcal{M}$ denotes the set of all possible messages that can be encrypted, and $m_1$ and $m_2$ are any two messages in the scheme. The operation $\star$ can be any binary operation (e.g. addition or multiplication).  

\begin{mybox}[boxsep=0pt,
	boxrule=1pt,
	left=4pt,
	right=4pt,
 	top=4pt,
 	bottom=4pt,
	]

 \textbf{Tech Tips:}
Homomorphic encryption safeguards privacy  during the inference stage by encrypting both the model parameters and input data. With HE, computations can be performed directly on encrypted data, allowing the model to make predictions without decrypting sensitive information. This process ensures that neither the raw data nor the model architecture is exposed in their unencrypted form, preserving privacy throughout the inference process. Decryption of the results is only done by trusted parties possessing the decryption key, maintaining the confidentiality of the information. Additionally, HE facilitates secure outsourcing of computations to untrusted servers, enabling organizations to utilize external resources without compromising data privacy. \looseness=-1
\end{mybox}

 We now introduce privacy inference approaches based on HE \cite{chen2022x,hao2022iron,lu2023bumblebee,zimerman2023converting,liu2023llms}. 
The THE-X \cite{chen2022x} presented a novel approach for enabling privacy-preserving inference on pre-trained transformer models using homomorphic encryption, in which they utilized ReLU to replace GELU and used approximation methods for SoftMax and LayerNorm to support the fully HE operations. However, THE-X may lead to privacy leakages as it poses intermediate results to the client during the computing of ReLU. 
Iron \cite{hao2022iron} focused on enhancing privacy in client-server settings, where clients have private inputs and servers hold proprietary models. It introduces several new homomorphic encryption-based protocols for matrix multiplication and complex non-linear functions (like Softmax, GELU activations, and LayerNorm) which are crucial in Transformer-based models. 
Bumblebee \cite{lu2023bumblebee} optimized homomorphic encryption-based protocols for large matrix multiplication and efficient, accurate protocols for non-linear activation functions in transformers, enhancing data privacy during inference. 
Zimerman \textit{et al.} \cite{zimerman2023converting} explored secure transformer models tailored for HE, which converts the operators to their polynomial equivalent. 
Liu \textit{et al.} \cite{liu2023llms} proposed a framework to enhance the efficiency of private inference on transformer-based models. It focuses on replacing computation-intensive operators (e.g., ReLU, GELU) in transformers with privacy-computing-friendly alternatives. The framework achieves significant reductions in private inference time and communication overhead while maintaining near-identical model accuracy.

 \subsubsection{\textbf{Multi-Party Computation}}
 Multi-Party Computation \cite{goldreich1998secure,NDSS23dong} is a cryptographic protocol that enables allows multiple parties (often mutually distrusting) to collaboratively perform a computation task while keeping their individual data private. This means that even though the parties are working together to compute a result, none of them can see the other parties' private data. The objective of secure multi-party Computation is to construct a secure protocol that allows multiple mistrustful participants to jointly compute a target function on their private inputs, while ensuring the accuracy of the output, and protecting and controlling their private inputs even in the presence of dishonest behavior. SMPC can be formally described as follows: 
Consider $n$ parties, denoted as $P_1,P_2,...,P_n$. Each party $P_i$ holds a private input $X_i$. There is a predefined function $f$ that takes $n$ inputs. This function is of the form $f:(X_1,X_2,...,X_n) \rightarrow Y$, where $X_i$ represents the input for party $P_i$ and $Y$ is the output using the secret data of all parties. Then, the parties compute the result $Y = (Y_1,Y_2,..,Y_n)$ based on the function $f(X_1,X_2,...,X_n)$ such that each party learns $Y$ (or a portion of $Y$ relevant to them) but learns nothing about the inputs $X_i$ of the other parties, for all $j \neq i$.

\begin{mybox}[boxsep=0pt,
	boxrule=1pt,
	left=4pt,
	right=4pt,
 	top=4pt,
 	bottom=4pt,
	]

 \textbf{Tech Tips:}
MPC enables secure aggregation of model updates in federated learning setups, allowing parties to collaboratively train a shared model. MPC ensures privacy during model inference by performing computations on encrypted data, shielding sensitive information from central servers.  MPC facilitates secure data labeling by allowing multiple parties to label data collaboratively without exposing raw labels, thus maintaining the confidentiality of sensitive information throughout the process.  

\end{mybox}
 
 Similar to HE, MPC is another crucial method that can be used to protect model privacy \cite{wang2022characterization,hou2023ciphergpt,ding2023east,akimoto2023privformer,dong2023puma}. 
Wang \textit{et al.} \cite{wang2022characterization} focused on the challenges and solutions for private inference in transformer models using MPC. While it advances the field of privacy-preserving inference, the complexity of MPC might affect practicality and efficiency. 
Hou \textit{et al.} \cite{hou2023ciphergpt} presented a framework CipherGPT for secure GPT model inference in a two-party setting. It introduces optimized cryptographic protocols for operations like matrix multiplication and GELU activation, essential for GPT models. The framework focuses on preserving privacy while ensuring the efficiency of the inference process. However, the specific focus on two-party settings may limit the framework's applicability in more diverse operational environments. 
Ding \textit{et al.} \cite{ding2023east} proposed a communication-efficient protocol called East for activation functions like GELU and tanh, as well as optimized protocols for softmax and Layer Normalization (LN). These protocols are designed to enhance performance by reducing runtime and communication overhead, ensuring the security of the scheme. 
Akimoto \textit{et al.} \cite{akimoto2023privformer} presented a MPC-based approach to secure inference of Transformer models in natural language processing using ReLU functions. This method addresses the challenge of computing the Transformer's attention mechanism efficiently and securely in an MPC setting. 
Dong \textit{et al.} \cite{dong2023puma} introduced PUMA, a framework for efficient and secure inference on Transformer models using replicated secret sharing. PUMA offers approximations for expensive non-linear functions (e.g., GeLU and softmax), which can also evaluate the large models like LLaMA-7B efficiently under MPC.

 \subsubsection{\textbf{Functional Secret Sharing}}
Function Secret Sharing (FSS) \cite{boyle2015function} involves dividing an original secret into multiple shares using a mathematical function (such as a polynomial), encoding the secret into each share in such a way that each is independent and insufficient to reveal the entire secret. These parts are then distributed to different participants, who can independently execute predetermined functions, such as arithmetic or logical operations, on their portion of the secret. These computations are carried out on secret shares that are in an encrypted or hidden state, preventing participants from obtaining any information about the original secret from their share alone. The results obtained by each participant are then aggregated, and when a sufficient number of shares are combined and computed, the outcome of executing the function on the entire secret is recovered. The security of this process lies in the fact that each share does not contain enough information to reveal the secret by itself; hence, even if some shares are compromised or participants are dishonest, the secret remains secure. The original secret's information is only revealed when the predetermined threshold is reached, that is, when a certain number of shares are correctly combined. 

% Generally, FSS satisfies the following properties:

% \begin{itemize}
%     \item Correctness: This property ensures that each evaluator (or participant) can correctly evaluate their shares of a secret function on an input.
    
%     \item Privacy: Privacy ensures that any individual secret share reveals no information about the function itself.

%     \item Efficiency: FSS ensures that the secret shares are succinct, meaning they are small in size, ideally sublinear in the size of the truth table for the original function.
% \end{itemize}

 \begin{mybox}[boxsep=0pt,
	boxrule=1pt,
	left=4pt,
	right=4pt,
 	top=4pt,
 	bottom=4pt,
	]

 \textbf{Tech Tips:}
In FSS, the LLM or function is partitioned into shares using cryptographic methods, with each party holding a share. During computation, parties perform operations on their shares using their private data, ensuring that individual inputs remain undisclosed. After computation, the parties collaboratively combine their shares to reconstruct the result of the function, maintaining privacy while revealing the final output.  \looseness=-1
\end{mybox}

As far as we know, there has been only one secure privacy inference approach based on Function Secret Sharing (FSS), which was proposed by Gupta \textit{et al.} \cite{gupta2023sigma}. The approach discussed a system named SIGMA for secure inference of transformer-based models, specifically focusing on Generative Pre-trained Transformers. SIGMA is designed to be efficient in terms of latency and communication overhead while maintaining standard 2-party computation (2PC) security by leveraging FSS. It introduces new FSS-based protocols for complex machine learning functionalities like Softmax and GeLU and optimizes them for GPU acceleration. SIGMA claims significant improvements in latency over state-of-the-art systems and demonstrates scalability to large GPT models. However, the paper does not explicitly outline specific disadvantages, which typically in such systems could include complexity of implementation, computational resource requirements, or potential limitations in the types of models or data that can be securely processed.

 \subsubsection{\textbf{Differential Privacy in Inference}}
% about inference???
Similarly, differential privacy can also be applied in the inference stage of LLM, providing a crucial layer of privacy protection during the generation of model predictions or outputs.

 \begin{mybox}[boxsep=0pt,
	boxrule=1pt,
	left=4pt,
	right=4pt,
 	top=4pt,
 	bottom=4pt,
	]

 \textbf{Tech Tips:}
In the inference stages of LLMs, DP can introduce noise to model outputs to safeguard individual data privacy while preserving prediction accuracy. Parameters are adjusted to manage the privacy budget effectively, with continuous monitoring ensuring a balance between privacy and utility over time.
\end{mybox}

Majmudar \textit{et al.} \cite{majmudar2022differentially} presented a method for ensuring differential privacy in the decoding process of LLMs. This approach aims to protect privacy during text generation. 
% fine-tuning and inference???
Du \textit{et al.} \cite{du2023dp} proposed a method for fine-tuning and inference in language models while maintaining differential privacy during the forward pass. It tackles the challenge of protecting privacy during both fine-tuning and inference phases.  
% inference
Mai \textit{et al.} \cite{mai2023split} introduced the Split-and-Denoise method, combining local differential privacy with a denoising technique to protect privacy in large language model inference. 
% Also, in addition to above approaches, there are other approaches that have been proposed to protect model privacy, but do not adopt HE, MPC or FSS \cite{wang2023privatelora,zhou2022textfusion,yuan2023secure}. 
%
Zhou \textit{et al.} \cite{zhou2022textfusion} introduced a method for privacy-preserving inference in pre-trained models using token fusion. The advantage is maintaining privacy during inference, but it could impact the inference accuracy or efficiency. 
Yuan \textit{et al.} \cite{yuan2023secure} detailed a three-party protocol for secure Transformer model inference, safeguarding both model parameters and user data. It applies permutation instead of complex encryption, offering strong security with practical feasibility for global matrix multiplication-based layers.

\begin{tcolorbox}[colback=blue!2!white,colframe=blue!20!black,title=Finding: Cryptography-based Private Inference]
    Privacy protection techniques grounded in Homomorphic Encryption (HE), Multi-Party Computation (MPC), and Functional Secret Sharing (FSS) offer demonstrable security assurances within rigorously defined threat models, as indicated in Table~\ref{table: performance comparison}. Nevertheless, limitations in performance and efficiency present obstacles to their near-term adoption by prominent model service providers. Even though these techniques have enhanced the efficiency of critical components, their experimental results demonstrate that deploying HE, MPC, and FSS might lead to degraded performance. Alternative approaches often rely on principles of obfuscation, yet their levels of randomness and security are weaker than cryptography-based solutions, and they typically consider specific attacks.
\end{tcolorbox}

% \subsection{Privacy Protection for LLM Outputs}
\subsection{Detection-based Approaches}
In existing research on Language Models (LMs), some efforts focuses on detecting privacy leaks\cite{lukas2023analyzing, brown2022simple,wu2023depn,shvartzshnaider2019vaccine}. These studies predominantly examine whether the content generated by LMs directly exposes data privacy or if such privacy can be inferred through contextual associations. This approach is equally applicable to LLMs, suggesting a viable pathway for assessing and mitigating privacy risks in more advanced linguistic computational models.

\begin{mybox}[boxsep=0pt,
	boxrule=1pt,
	left=4pt,
	right=4pt,
 	top=4pt,
 	bottom=4pt,
	]

 \textbf{Tech Tips:}
 Detection-based methods for protecting the privacy of LLM involve identifying and mitigating potential privacy risks in the text generated by these models which two main strategies: (i) Direct detection methods involve directly examining the text generated by LLMs to identify privacy leaks. (ii)
Contextual inference detection methods focus on identifying privacy breaches that may not be explicitly evident in the generated text but can be deduced through contextual correlations. 

\end{mybox}
% Directly detecting the text generated by LLMs for privacy leak risks represents the most direct approach, providing clear and immediate advantages in pinpointing potential data exposures. Text data generated by LLMs may not explicitly reveal private information, yet it is possible for attackers to deduce sensitive details through contextual correlations. Therefore, the identification of contextually related privacy breaches is of paramount importance.
 \subsubsection{\textbf{Direct Detection}}  Kim \textit{et al.}\cite{kim2023propile} developed a black-box probing method to evaluate privacy risks in LLMs by using crafted prompts to elicit Personally Identifiable Information (PII) from model outputs. This approach assesses the likelihood of LLMs inadvertently revealing PII, offering a targeted strategy for understanding privacy vulnerabilities in generated text. Phute \textit{et al.}\cite{phute2023llm} unveiled a zero-shot defense strategy for LLMs aimed at curbing harmful content generation. By deploying a harm classifier from the same LLM, this method significantly reduces the efficacy of adversarial attacks, eliminating the need for fine-tuning. Chen \textit{et al.} \cite{chen2023jailbreaker} developed a moving target defense system for LLMs to counter adversarial attacks, using N-Gram models and naive Bayes classification for evaluating responses and BERT for assessing question-answer coherence, effectively distinguishing between beneficial and malicious content.

\subsubsection{\textbf{Contextual Inference Detection}} 
Mireshghallah \textit{et al.}\cite{mireshghallah2023llms} introduced CONFAIDE, a benchmark that evaluates LLMs' privacy reasoning across four complexity levels, revealing notable deficiencies in models like GPT-4 and ChatGPT in terms of privacy preservation and social reasoning. Huang \textit{et al.} \cite{huang2022large} proposed a framework to assess PLMs' risk of privacy leakage, focusing on email addresses. Their approach, which analyzes memorization and association, highlights vulnerabilities in how models might unintentionally disclose or link email addresses to individuals.

 \begin{tcolorbox}[colback=blue!2!white,colframe=blue!20!black,title=Finding: Detection-based Approaches]
    Due to the inherent complexity and variability of text data, scrutinizing the outputs of LLMs in practical applications has its limitations. Attackers can exploit these limitations by crafting impermissible outputs from seemingly permissible ones\cite{glukhov2023llm}. This underscores the necessity for advanced and dynamic security measures, beyond simple output filtering or static rules, to effectively counteract sophisticated manipulation techniques and ensure the integrity and safety of LLMs applications.
\end{tcolorbox}

\subsection{Hardware-based Approaches}
Hardware-based approaches for protecting the privacy of LLM focus on leveraging specialized hardware features and technologies to establish secure execution environments and safeguard data during processing.\looseness=-1

\begin{mybox}[boxsep=0pt,
	boxrule=1pt,
	left=4pt,
	right=4pt,
 	top=4pt,
 	bottom=4pt,
	]

 \textbf{Tech Tips:} 
 Hardware-based Approaches such as Trusted Execution Environments (TEEs), hardware virtualization, secure enclaves, hardware Root of Trust (RoT), and encrypted processing, aim to ensure the confidentiality, integrity, and privacy of both the model parameters and the data being processed. 
\end{mybox}

\subsubsection{\textbf{Data Locality}}
PrivateLoRA \cite{wang2023privatelora} leveraged edge devices' storage for private data and personalized parameters, while utilizing the cloud for computational enhancement. It splits model parameters across the cloud and edge devices and transmits only unreadable activations and gradients to maintain data locality. The method integrates three sequential low-rank matrices for weight adaptation and reduces communication overhead through Low Rank Residual Transmission. It ensures data locality by keeping personalized parameters on edge devices and raw data derivatives on the cloud. The model targets query, key, and value projections in self-attention for adaptation to minimize communication overhead. PrivateLoRA is a paradigm that powers a heterogeneously distributed inference and training cycle, achieving high throughput and performance on smart phones.

 \subsubsection{\textbf{Confidential Computing with Trusted Execution Environment (TEE)}}
Confidential computing aims to address this gap by safeguarding data even while it is being processed. One key technology used in confidential computing is Trusted Execution Environments (TEEs). A TEE is a secure area of a computer's processor that ensures code and data loaded inside it are protected from unauthorized access or modification, even from the operating system or hypervisor. TEEs provide a secure environment where sensitive computations can be performed, ensuring the confidentiality and integrity of the data being processed~\cite{chen2023verified, zhu2020enabling, 9705115, li2022sok,luo2022security,weng2022peripheral,liu2020manually,pearson2020sic}.

The NVIDIA H100 GPU, featuring support for confidential computing, enhances data privacy by establishing a secure execution environment through hardware virtualization and a TEE \cite{dhanuskodi2023creating}. This environment ensures that data and code are processed securely during training or inference, preventing unauthorized access or modification by unauthorized users. By anchoring security measures in an on-die hardware root of trust (RoT), NVIDIA ensures the integrity of the GPU's boot sequence and establishes a chain of trust through cryptographic attestation. Furthermore, NVIDIA continues to enhance the security and integrity by incorporating features such as encrypted firmware, firmware revocation, and fault injection countermeasures. The TEEs applied in \cite{southsecure} protect privacy by securely executing custodial operations, encrypting and controlling access to data, and facilitating encrypted transmission of user queries and prompts.  Huang et al.~\cite{huang2024fast} introduced a method deploying TEEs on both client and server sides, implementing secure communication and split fine-tuning of a language model to maintain accuracy.

%In summary, various research efforts were proposed to improve privacy and security in model inference. Some focus on private inference using complex computational techniques, while others offer frameworks for large model inference with enhanced data privacy, albeit with added computational overhead. The challenges and solutions for using MPC are also discussed, highlighting the trade-offs between privacy preservation, practicality, and efficiency. Additionally, approaches using homomorphic encryption and function secret sharing are mentioned, aiming to secure inference with a focus on maintaining data privacy, albeit potentially increasing latency and computational demands. Overall, these research efforts strive for a balance between privacy, security, and efficiency in transformer model inference.

\section{Challenges and Future Directions}

\subsection{Difficulties in Understanding Black-Box LLMs}  

\subsubsection{Challenges} Pre-trained LLMs are often treated as black box models \cite{hong2023cyclealign, wang2023augmenting}, meaning that their internal workings and decision-making processes are not fully transparent or interpretable. This opacity makes it challenging to analyze and understand how these models handle sensitive information and whether they inadvertently leak privacy. In addition, LLMs are trained on vast amounts of diverse data, which may include sensitive or personally identifiable information. Understanding how these models process and retain such data without compromising privacy is inherently complex, especially given the intricate relationships between input data and model outputs. Language is dynamic and context-dependent, leading to challenges in predicting how LLMs will behave in various real-world scenarios. Privacy risks may vary depending on the context in which the model is deployed, making it difficult to generalize findings across different applications or domains. 

\subsubsection{Future Directions} Developing techniques to interpret and explain the decisions of pre-trained LLMs can shed light on their privacy implications. This may involve analyzing model activations, attention mechanisms, or other internal representations to identify potential privacy vulnerabilities. 
Conducting adversarial testing to evaluate the robustness of pre-trained LLMs against privacy attacks. For example, adversarial examples can be generated to probe the model's behavior and identify weaknesses that may lead to privacy breaches \cite{chao2023jailbreaking}. 
Besides, we can focus on developing fine-tuning techniques that explicitly consider privacy concerns, such as differential privacy-aware optimization or adversarial training with privacy objectives. These techniques aim to mitigate privacy risks during the fine-tuning process.

\subsection{Privacy in Multimodal LLMs}
\subsubsection{Challenges}  The majority of research on LLMs has focused on purely textual models such as GPT and BERT. As a result, there may be a tendency for researchers to prioritize investigating the privacy implications of these models, leaving less attention on Multimodal LLMs. Multimodal LLMs, which integrate both textual and visual information, are a relatively recent development compared to their purely textual counterparts \cite{li2023seed, mesko2023impact}. As such, there has been less time for researchers to explore and investigate their privacy implications thoroughly. Multimodal LLMs process a more diverse range of data types, including text, images, and possibly other modalities such as audio or video. Analyzing the privacy implications of such complex and heterogeneous data poses additional challenges compared to purely textual data, which may deter some researchers from delving into this area. 

\subsubsection{Future Directions}  Redefining privacy in Multimodal LLMs is necessary to address the increased data complexity, unique privacy risks, intermodal interactions, user expectations, and regulatory considerations associated with multimodal data processing. Developing techniques to fuse different modalities while preserving user privacy is an important research direction. This could involve exploring encryption methods, differential privacy techniques, or novel privacy-preserving machine learning algorithms tailored to multimodal data. Conducting adversarial analyses to identify potential vulnerabilities and privacy risks in Multimodal LLMs. This could involve exploring adversarial attacks and defenses specific to multimodal data, such as perturbing images or textual inputs to compromise privacy.

\subsection{Privacy in Personalized LLMs}
\subsubsection{Challenges} Personalized LLMs may store and process sensitive user data, such as personal conversations, search queries, or browsing history. If not adequately protected, this data could be vulnerable to unauthorized access or misuse, leading to privacy breaches and potential harm to individuals. Personalized LLMs have the capacity to infer personal information about users based on their interactions with the model. This includes sensitive attributes such as health status, political views, financial situation, or intimate preferences. Such inferences could be unintentionally revealed through model responses or recommendations, compromising user privacy. Numerous small-scale enterprises offer users specialized large-scale model services tailored to vertical domains, encompassing sectors such as judiciary, education, and finance. These expansive models entail a greater incorporation of domain-specific personal data. However, owing to the comparatively limited privacy safeguarding capabilities inherent in small-scale enterprises, the susceptibility to user privacy breaches is heightened, potentially precipitating irreversible ramifications.

\subsubsection{Future Directions}
To safeguard personalized fine-tuning of LLMs from privacy leakage, we need to explore architectures specifically designed with privacy \cite{huang2023firewallm}. In addition, we can develop a combination of techniques. This includes implementing differential privacy methods to add noise during training, utilizing federated learning to train models locally on user devices, employing secure multi-party computation to jointly train models without sharing private data directly, introducing data perturbation to prevent memorization of sensitive information. We can also apply regularization methods to prevent overfitting, and exploring privacy-preserving architectures designed specifically for protecting sensitive data during fine-tuning.\looseness=-1
 
\subsection{Privacy Protection Throughout the Entire Creation Process of LLMs}
\subsubsection{Challenges} Given the intricate complexity involved in training LLMs, privacy protection research tends to dissect various phases of LLM development and deployment, including pre-training, prompt tuning, and inference. Nevertheless, each segment within the LLM lifecycle harbors its own set of privacy vulnerabilities, and these stages do not operate in isolation \cite{evertz2024whispers}. For instance, privacy breaches detected during the inference phase might originate from potential backdoors introduced during pre-training. Thus, safeguarding privacy comprehensively across large models demands concurrent scrutiny of multiple stages, a task that also introduces complexities and challenges into privacy protection efforts.

\subsubsection{Future Directions} Protecting the privacy of LLMs throughout their creation process is paramount and requires a multifaceted approach. Firstly, during data collection, minimizing the collection of sensitive information and obtaining informed consent from users are critical steps. Data should be anonymized or pseudonymized to mitigate re-identification risks. Secondly, in data preprocessing and model training, techniques such as federated learning, secure multiparty computation, and differential privacy can be employed to train LLMs on decentralized data sources while preserving individual privacy. Additionally, conducting privacy impact assessments and adversarial testing during model evaluation ensures potential privacy risks are identified and addressed before deployment.
In the deployment phase, privacy-preserving APIs and access controls can limit access to LLMs, while transparency and accountability measures foster trust with users by providing insight into data handling practices. Ongoing monitoring and maintenance, including continuous monitoring for privacy breaches and regular privacy audits, are essential to ensure compliance with privacy regulations and the effectiveness of privacy safeguards. By implementing these measures comprehensively throughout the LLM creation process, developers can mitigate privacy risks and build trust with users, thereby leveraging the capabilities of LLMs while safeguarding individual privacy.

\subsection{Hardware-assisted Privacy Protection} 
\subsubsection{Future Directions} NVIDIA Confidential Computing provides a comprehensive suite of privacy-enhancing features and technologies that safeguard LLM data and operations against unauthorized access, manipulation, and breaches, thereby ensuring the confidentiality and integrity of sensitive information throughout the LLM lifecycle. In the future, we can integrate confidential computing capabilities into LLM workflows, ensuring comprehensive privacy protection across the entire lifecycle, while continued innovation in GPU security features, such as encrypted firmware and fault injection countermeasures, reinforces the company's commitment to advancing data privacy safeguards for sensitive workloads.

\section{Conclusion}
In this paper, we thoroughly investigates the data privacy concerns associated with LLMs, focusing on privacy leakage, privacy attacks, and the pivotal technologies for privacy protection during various stages of LLM privacy inference, including federated learning, differential privacy, knowledge unlearning, and hardware-assisted privacy protection. By conducting a detailed analysis of the strengths and weaknesses of existing approaches, this study highlights the challenges and limitations in LLMs and proposes directions for future work. This research is of significant importance for deepening our understanding of data privacy issues in LLMs and promoting further exploration and improvement in LLMs. 
% \section{Acknowledgment}

\ifCLASSOPTIONcaptionsoff
	\newpage
\fi

\bibliographystyle{IEEEtran}
\bibliography{references}

\end{document}